\documentclass[prb,citeautoscript,twocolumn]{revtex4}%
\usepackage{amsfonts}
\usepackage{amsmath}
\usepackage{graphicx}
\usepackage{dcolumn}
\usepackage{bm}
\usepackage{amssymb}%
\newcommand{\bml}{\begin{mathletters}}
\newcommand{\eml}{\end{mathletters} \hspace{-5pt}}
\allowdisplaybreaks[1]
\begin{document}
\title{Two-channel Anderson Impurity Model:\\Single-electron Green's function, self-energies, and resistivity}
\author{Henrik~Johannesson}
\affiliation{Institute of Theoretical Physics, Chalmers University of Technology and
G\"{o}teborg University, SE-412 96 G\"{o}teborg, Sweden}
\author{C.~J.~Bolech}
\affiliation{Universit\'{e} de Gen\`{e}ve, DPMC, 24 Quai Ernest Ansermet, CH-1211
Gen\`{e}ve 4, Switzerland}
\author{Natan~Andrei}
\affiliation{Center for Materials Theory, Serin Physics Laboratory, Rutgers University,
Piscataway, New Jersey 08854-8019, USA}

\begin{abstract}
We compute exactly the low-energy single-electron Green's function, the
impurity and electron self-energies, and the resistivity for the two-channel
Anderson impurity model. These results are obtained by exploiting the boundary
conformal field theory identified from the Bethe Ansatz solution of the model.
Using that solution we can make contact with the parameters of the original
Hamiltonian and provide the detailed crossover between the two integer valence
limits. Our results generalize those obtained previously in the context of the
two-channel Kondo model.

\end{abstract}
\date{November 2004}
\maketitle

\section{Introduction\label{section1}}

Multi-channel impurity physics is a well-established route to non-Fermi-liquid
behavior. Its appeal is therefore understandable as an essential ingredient in
many of the various theories that try to explain the unusual characteristics
of numerous systems ranging from heavy fermions \cite{Hewson} to mesoscopic
point contacts \cite{Kouwenhoven}. In this context, the two-channel Anderson
impurity model occupies a central place. It was first introduced by Cox in an
attempt to model the physics of certain \textrm{U}-based heavy fermions
\cite{Cox}, among which the compound \textrm{UBe}$_{13}$ is a prime example.
Following that line of thought, a dilute concentration of uranium in a
\textrm{ThBe}$_{13}$ matrix will correspond to a metallic system with a feeble
concentration of two-channel impurity centers. Impurity corrections to the
different transport properties of that system should display fractional power
laws indicative of non-Fermi-liquid behavior. For instance, corrections to the
resistivity would display $\sqrt{T}$ dependence at low temperatures
\cite{Aliev} and should constitute a particular experimentally accessible observable.

Indeed, transport measurements are good candidates for experiments, since they
can also be performed in mesoscopic systems for which bulk thermodynamic
measurements, possible for heavy fermions and other materials, are
ineffectual. In this context, the two-channel Anderson model was already used
successfully as the starting point of Non-Crossing Approximation (NCA)
calculations to model the temperature dependence of the differential
conductance of \textrm{Cu} point contacts \cite{Hettler,Hettler2}.

The two-channel Anderson impurity model \cite{CoxZawadowski,SchillerAndersCox,
KrohaWolfle} describes the interaction of three-dimensional electrons with a
local impurity carrying both spin and quadrupolar degrees of freedom. These
degrees of freedom correspond to the lowest-energy configurations of a uranium
impurity in charge states $\mathrm{U}^{4+}$\ ($5f^{2}$) and $\mathrm{U}^{3+}%
$\ ($5f^{3}$). Taking into account spin-orbit coupling and crystal-field
splitting in a cubic background one ends up considering the minimal scenario
of a (quadrupolar) $\Gamma_{3}$ doublet representation of the cubic group as
the lowest-energy multiplet in $\mathrm{U}^{4+}$ and a Kramers doublet
$\Gamma_{6}$ as the lowest-energy configuration in $\mathrm{U}^{3+}$. These
levels hybridize with those electrons from the conduction band effectuating
the transition from one doublet to the other. Starting from the full electron
field $\Psi(\vec{x})$, one carries out an expansion in harmonics corresponding
to cubic symmetry, retaining only $\Psi_{\alpha\sigma}%
^{\mathop{\phantom{\dagger}}}(r)$ which has the appropriate symmetry to couple
to the impurity, with $\alpha=\pm$ denoting the quadrupolar degrees of freedom
and $\sigma=\uparrow,\downarrow$ denoting magnetic (spin) degrees of freedom.

We then proceed to write the field in terms of 1D right (left) moving fields
$\psi_{R\alpha\sigma}^{\mathop{\phantom{\dagger}}}$ ($\psi_{L\alpha\sigma
}^{\mathop{\phantom{\dagger}}}$), representing the incoming (outgoing) radial
components of the 3D electron fields that couple to the impurity
\cite{CoxZawadowski},
\begin{equation}
\Psi_{\alpha\sigma}^{\mathop{\phantom{\dagger}}}(r)=\frac{i}{2\sqrt{2}\pi
r}\left[  \mbox{e}^{-ik_{F}r}\psi_{L\alpha\sigma}^{\mathop{\phantom{\dagger}}}%
(r)-\mbox{e}^{ik_{F}r}\psi_{R\alpha\sigma}^{\mathop{\phantom{\dagger}}}%
(r)\right]  , \label{swaveprojection}%
\end{equation}
with a \textquotedblleft free electron\textquotedblright\ boundary condition
$\psi_{L\alpha\sigma}^{\mathop{\phantom{\dagger}}}(0)=\psi_{R\alpha\sigma
}^{\mathop{\phantom{\dagger}}}(0)$ imposed at the origin $r=0$ \cite{ALNP}.

The Hamiltonian is then given by
\begin{equation}
H=H_{\text{{\footnotesize bulk}}}+H_{\text{{\footnotesize ion}}}%
+H_{\text{{\footnotesize hybr}}}\,, \label{hamiltonian}%
\end{equation}
where
\begin{align}
H_{\text{{\footnotesize bulk}}}  &  =\int_{0}^{\infty}dr\left[  :\!\psi
_{L\alpha\sigma}^{\dagger}(r)(i\partial_{r})\psi_{L\alpha\sigma}%
^{\mathop{\phantom{\dagger}}}(r)\!:\right.  -\nonumber\\
&  \qquad\qquad\qquad-\left.  :\!\psi_{R\alpha\sigma}^{\dagger}(r)(i\partial
_{r})\psi_{R\alpha\sigma}^{\mathop{\phantom{\dagger}}}(r)\!:\right]
\label{bulk}\\
H_{\text{{\footnotesize ion}}}  &  =\epsilon_{s}f_{\sigma}^{\dagger}f_{\sigma
}^{\mathop{\phantom{\dagger}}}+\epsilon_{q}b_{\bar{\alpha}}^{\dagger}%
b_{\bar{\alpha}}^{\mathop{\phantom{\dagger}}}\label{ion}\\
H_{\text{{\footnotesize hybr}}}  &  =V[\psi_{L\alpha\sigma}^{\dagger
}(0)b_{{\bar{\alpha}}}^{\dagger}f_{\sigma}^{\mathop{\phantom{\dagger}}}%
+f_{\sigma}^{\dagger}b_{{\bar{\alpha}}}^{\mathop{\phantom{\dagger}}}%
\psi_{L\alpha\sigma}^{\mathop{\phantom{\dagger}}}(0)]\,. \label{hybr}%
\end{align}
The conduction electrons here hybridize with the impurity via a matrix element
$V$ \cite{Vfootnote}, with the impurity modeled by a quadrupolar (magnetic)
doublet of energy $\epsilon_{q}$ ($\epsilon_{s}$), created by a boson
(fermion) operator $b_{\bar{\alpha}}^{\dagger}$ ($f_{\sigma}^{\dagger}$).
\cite{footnote2} Strong Coulomb repulsion implies single occupancy of the
localized levels: $f_{\sigma}^{\dagger}f_{\sigma}^{\mathop{\phantom{\dagger}}}%
+b_{\bar{\alpha}}^{\dagger}b_{\bar{\alpha}}^{\mathop{\phantom{\dagger}}}=1$.
The free part of the Hamiltonian, $H_{\text{{\footnotesize bulk}}}$, defines a
linearized spectrum around the Fermi level. The Fermi velocity is set to
unity, with the resulting 1D density of states $\rho=1/(2\pi)$. Normal
ordering is taken with respect to the filled Fermi sea.

The model in Eq. (\ref{hamiltonian}) was recently solved by two of us using a
\emph{Bethe Ansatz} \cite{BolechAndrei}. A complete description of spectrum
and thermodynamics was given, and it was found that at low temperatures the
theory is attracted to a line of fixed points parametrized by the impurity
charge valence $n_{c}$ (where $n_{c}=\langle f_{\sigma}^{\dagger}f_{\sigma
}\rangle$ measures the average charge localized at the impurity site). Near
integral charge valence $n_{c}\simeq1\ $($n_{c}\simeq0$) a magnetic
(quadrupolar) moment forms at intermediate temperatures. This moment is then
screened by the conduction electrons as the temperature is lowered, leading to
a zero-temperature entropy $S_{0}^{\text{imp}}=k_{B}\ln\sqrt{2}$ and impurity
specific heat $C_{\text{v}}^{\text{imp}}\sim T\ln T$, typical of two-channel
Kondo physics. In the mixed-valence regime one finds the same low-temperature
behavior, but without the formation of a magnetic or quadrupolar moment at
intermediate temperatures.

In previous work we constructed the Boundary Conformal Field Theory (BCFT)
which describes the approach to criticality \cite{JAB}. The leading scaling
operators were identified---including the exactly marginal operator that
generates the line of fixed points---and all physical scales and BCFT
parameters were determined explicitly via a numerical fit to the exact
solution in Ref.~[\onlinecite{BolechAndrei}]. This allowed us to go beyond the
\emph{Bethe Ansatz} approach and derive the critical exponents of the Fermi
edge singularities caused by time-dependent hybridization between conduction
electrons and impurity. Our results challenged those obtained by more
conventional, approximate schemes (see, \textit{e.g.,} Ref.~[\onlinecite{CoxRuckenstein}]).

In the present work we take the BCFT formulation one step further and extract
the exact space- and time-dependent single-electron Green's function of the
model. This allows us to calculate the self-energies of the conduction
electrons and of the impurity, as well as the zero-temperature resistivity and
leading temperature-dependent term.

The analysis is most easily performed by generalizing that of the
multi-channel Kondo model in Refs.~[\onlinecite{GF1}] and [\onlinecite{GF2}].
In fact, the very structure of the Green's function, as well as that of the
leading terms of the resistivity, can be read off directly from the
corresponding result for the two-channel Kondo model \cite{GF1, GF2}. The only
essential new element in the analysis is how to properly introduce the scales
and amplitudes that determine the influence from the magnetic and quadrupolar
degrees of freedom as one moves away from the integer valence limits.

In the next section we combine results from Ref.~[\onlinecite{JAB}] and
Refs.~[\onlinecite{GF1, GF2}] to obtain the single-electron Green's function
of the model. In Sec.~III we use this result to derive the self-energy of the
impurity, and in Sec.~IV we report on the calculation of the resistivity.
Section V contains a summary and a discussion of our results.

\section{Green's Function}

At sufficiently low energies (or large distances) a quantum impurity
interacting with conformal (i.e., linear dispersion) electrons can be
represented by a conformally invariant boundary condition---this is the key
idea of the BCFT formulation of a quantum impurity problem
\cite{AffleckReview}. For the two-channel Anderson model, this boundary
condition [which supersedes the trivial \textquotedblleft
free-electron\textquotedblright\ boundary condition $\psi_{L\alpha\sigma
}^{\mathop{\phantom{\dagger}}}(0)=\psi_{R\alpha\sigma}%
^{\mathop{\phantom{\dagger}}}(0)$], is most easily described via a
\textquotedblleft gluing condition\textquotedblright\ on the charge, spin, and
flavor conformal towers that make up its spectrum (for details, see Ref.
[\onlinecite{JAB}]). The dependence on the new boundary condition
(\emph{alias} the impurity) is picked up by the time-ordered left-right
single-electron Green's functions
\begin{multline}
G_{LR}(\tau;r_{1},r_{2})=G_{RL}^{\ast}(\tau;r_{1},r_{2})\label{1DLR}\\
\equiv\langle\psi_{L\alpha\sigma}^{\mathop{\phantom{\dagger}}}(\tau,r_{1}%
)\psi_{R\alpha\sigma}^{\dagger}(0,r_{2})\rangle
\end{multline}
(with $\tau$ imaginary time) via the one-particle \textit{S}-matrix
\begin{equation}
S_{(1)}(\omega_{{\footnotesize F}})=\mbox{e}^{2i\delta_{{\footnotesize F}}%
}C_{(1)}(\omega_{{\footnotesize F}}). \label{S}%
\end{equation}
Here $C_{(1)}(\omega_{{\footnotesize F}})$ is the amplitude for a single
electron to scatter elastically off the impurity at the Fermi level
$\omega_{{\footnotesize F}}$, and $\delta_{{\footnotesize F}}=\delta
(\omega_{{\footnotesize F}})$ is the corresponding single-electron scattering
phase shift. At large (mean) distances from the boundary, $\mid\!r_{1}%
-r_{2}\!\mid\,\gg a$, with $a$ some characteristic microscopic scale, one
finds that \cite{GF1}
\begin{equation}
G_{LR}(\tau;r_{1},r_{2})\sim\frac{S_{(1)}(\omega_{F})}{\tau+i(r_{1}+r_{2})}.
\label{nonchiralGF}%
\end{equation}
Thus, at the level of the left-right Green's function, the boundary condition
that emulates the presence of the impurity is coded by $S_{\left(  1\right)
}(\omega_{{\footnotesize F}})$. In contrast, the large-distance left-left (LL)
and right-right (RR) Green's functions $G_{mm}(\tau;r_{1},r_{2})\equiv
\langle\psi_{m\alpha\sigma}^{\mathop{\phantom{\dagger}}}(\tau,r_{1}%
)\psi_{m\alpha\sigma}^{\dagger}(0,r_{2})\rangle$ with $m=L,R$ are insensitive
to the particular boundary condition imposed:\cite{GF1}
\begin{multline}
G_{LL}(\tau;r_{1},r_{2})=G_{RR}^{\ast}(\tau;r_{1},r_{2})\label{chiralGF}\\
\sim\frac{1}{\tau+i(r_{1}-r_{2})}~\text{,}\qquad\mid\!r_{1}-r_{2}\!\mid\,\gg
a\text{.}%
\end{multline}

Turning now to the the 3D electron field Green's function $\mathcal{G}%
(\tau,r_{1},r_{2})\equiv\langle\Psi_{\alpha\sigma}%
^{\mathop{\phantom{\dagger}}}(\tau,r_{1})\Psi_{\alpha\sigma}^{\dagger}%
(0,r_{2})\rangle$ and expressing it in terms of the 1D propagators in
Eqs.~(\ref{nonchiralGF}) and (\ref{chiralGF}) one obtains
%
\begin{align}
\mathcal{G}(\tau,r_{1},r_{2}) &  =\,\mbox{e}^{-ik_{F}(r_{1}-r_{2})}G_{LL}%
(\tau,r_{1},r_{2})\label{3DGF}\\
&  +~\mbox{e}^{ik_{F}(r_{1}-r_{2})}G_{RR}(\tau,r_{1},r_{2})\nonumber\\
&  +~\mbox{e}^{-ik_{F}(r_{1}+r_{2})}G_{LR}(\tau,r_{1},r_{2})\nonumber\\
&  +~\mbox{e}^{ik_{F}(r_{1}+r_{2})}G_{RL}(\tau,r_{1},r_{2}).\nonumber
\end{align}
Continuing the 1D electron fields analytically to the full line $-\infty
<r<\infty$ [with $\psi_{R\alpha\sigma}^{\mathop{\phantom{\dagger}}}%
(\tau,r)=\psi_{L\alpha\sigma}^{\mathop{\phantom{\dagger}}}(\tau,-r)$],
averaging over impurity locations (thus restoring translational invariance),
and exploiting a \textit{T}-matrix formulation \cite{Mahan}, the
Fourier-transformed Green's function in Eq.~(\ref{3DGF}) can be cast in the
standard form
\begin{equation}
\mathcal{G}(\omega_{n},k)=\frac{1}{i\omega_{n}-\epsilon_{k}-\Sigma(\omega
_{n})},\label{Green3D}%
\end{equation}
with the self-energy $\Sigma(\omega_{n})$ given by
\begin{equation}
\Sigma(\omega_{n})=-in_{i}\frac{1-\mbox{e}^{2i\delta_{F}}C_{1}(\omega
_{{\footnotesize F}})}{2\pi g_{{\footnotesize F}}}\mbox{sgn}(\omega
_{n}).\label{selfenergy}%
\end{equation}
Here $n_{i}$ is the density of a dilute random distribution of impurities,
$g_{{\footnotesize F}}$ is the 3D \textquotedblleft
free-electron\textquotedblright\ density of states at the Fermi level, and
$\operatorname{sgn}(\omega_{n})$ is the sign function, with $\omega_{n}%
=2\pi(n+1/2)k_{B}T$,$~n=0,\pm1,\pm2,\ldots$, Matsubara frequencies. In order
to tie the self-energy $\Sigma(\omega_{n})$ to the two-channel Anderson model
in Eq.~(\ref{hamiltonian}) we need to determine the \textit{S}-matrix in Eq.
(\ref{S}) that is associated with the hybridization interaction in
Eq.~(\ref{hybr}). The amplitude $C_{(1)}(\omega_{{\footnotesize F}})$ can be
determined in exact analogy with the two-channel Kondo problem in
Ref.~[\onlinecite{GF1}]. Within the BCFT formalism $C_{1}(\omega
_{{\footnotesize F}})$ gets expressed as a certain combination of so-called
\textquotedblleft modular \textit{S} matrices\textquotedblright%
\cite{DiFrancesco}
\begin{equation}
S_{j_{f}}^{j_{f}^{\prime}}=\frac{1}{\sqrt{2}}\mbox{sin}[\pi(2j_{f}%
+1)(2j_{f}^{\prime}+1)/4],\label{Smodular}%
\end{equation}
with structure and allowed values of the quantum numbers $j_{f},j_{f}^{\prime
}=0,1/2,1$ determined by the $\mathrm{SU}(2)_{2}$ Kac-Moody symmetry of the
flavor sector \cite{JAB}. Specifically,
\begin{equation}
C_{1}(\omega_{{\footnotesize F}})=\frac{S_{1/2}^{1/2}S_{0}^{0}}{S_{1/2}%
^{0}S_{0}^{1/2}},\label{modularcombi}%
\end{equation}
and it follows from Eq.~(\ref{Smodular}) that
\begin{equation}
C_{1}(\omega_{{\footnotesize F}})=0.\label{zeroamplitude}%
\end{equation}
Thus, the outgoing scattering state has no single-electron component after
interaction with the impurity. This extreme non-Fermi-liquid behavior is the
same as for the two-channel spin Kondo model \cite{addFootnote} ($n_{c}=1$
limit of the two-channel Anderson model) and \emph{is not modified as one
moves into the mixed valence regime with} $n_{c}\neq0,1$. As seen from
Eq.~(\ref{S}), the impurity valence $n_{c}$, connected to the phase shift
$\delta_{{\footnotesize F}}$ via the Friedel-Langreth sum rule
\cite{Langreth}
\begin{equation}
\delta_{{\footnotesize F}}=\frac{\pi}{4}n_{c},\label{FriedelLangreth}%
\end{equation}
could only influence the scattering if there were a finite single-electron
cross section at the Fermi level. However, as $C_{1}(\omega_{{\footnotesize F}%
})=0$ independent of $n_{c}$, this does not happen. [Note that the phase shift
$\delta_{{\footnotesize F}}$ is that of an electron with spin $\sigma$
\emph{and} flavor index $\alpha$, hence the unconventional factor of $1/4$ in
Eq.~(\ref{FriedelLangreth}).\cite{JAB}]

To summarize the analysis thus far: The zero-temperature single-electron
Green's function is given by Eq.~(\ref{3DGF}), with
\begin{equation}
\Sigma(\omega_{n})=-in_{i}\frac{1}{2\pi g_{{\footnotesize F}}}%
\mbox{sgn}(\omega_{n}), \label{selfenergy2}%
\end{equation}
where $n_{i}$ and $g_{F}$ are defined after Eq.~(\ref{selfenergy}). We should
here stress that impurity-impurity interactions have been neglected in
Eq.~(\ref{selfenergy2}). As discussed in Ref.~[\onlinecite{GF1}] this type of
analysis is applicable only at temperatures high enough so that any remnant
effects from interimpurity interactions are washed out (but low enough so that
the theory is critical and the BCFT formulation remains valid). Also note that
the expressions for the 1D propagators in Eqs.~(\ref{nonchiralGF}) and
(\ref{chiralGF}) pick up corrections when $\mid\!r_{1}-r_{2}\!\mid\,\leq a$
\cite{MEJ}, implying that the result in Eqs.~(\ref{Green3D}) and
(\ref{selfenergy2}) gets modified when probing boundary correlations with
large momenta.

To obtain the leading finite-temperature and frequency corrections to
Eq.~(\ref{selfenergy2}) we need to consider the theory slightly off the line
of boundary fixed points. The scaling Hamiltonian
$H_{\text{{\footnotesize scaling}}}$ that governs the critical behavior close
to the fixed line was identified in Ref.~[\onlinecite{JAB}] as
\begin{multline}
H_{scaling}=H^{\ast}+\lambda_{c}J(0)+\lambda_{s}\mathcal{O}^{(s)}%
(0)+\lambda_{f}\mathcal{O}^{(f)}(0)\label{ScalingHamiltonian}\\
+\mbox{subleading terms}\,\text{.}%
\end{multline}
Here $H^{\ast}$ is the critical Hamiltonian that represents $H_{bulk}$ in
Eq.~(\ref{hamiltonian}), subject to the boundary condition that emulates the
impurity terms $H_{\text{{\footnotesize ion}}}$ and
$H_{\text{{\footnotesize hybr}}}$ in Eqs.~(\ref{ion}) and (\ref{hybr}),
respectively. These terms, which break particle-hole symmetry, also give rise
to the exactly marginal term $\lambda_{c}J(0)$ in
Eq.~(\ref{ScalingHamiltonian}), with $J(0)=\sum_{\alpha,\sigma}:\!\psi
_{\alpha\sigma}^{\dagger}\psi_{\alpha\sigma}^{\mathop{\phantom{\dagger}}}%
(0)\!:$ being the charge current at the impurity site and with $\lambda_{c}$
its conjugate scaling field. This is the operator that generates the line of
stable fixed points. \emph{Off} the fixed line the terms
$H_{\text{{\footnotesize ion}}}$ and $H_{\text{{\footnotesize hybr}}}$ allow
for additional \emph{irrelevant} boundary operators to enter the stage. Of
these, $\lambda_{s}\mathcal{O}^{(s)}(0)$ and $\lambda_{f}\mathcal{O}^{(f)}%
(0)$, both of scaling dimension $\Delta=3/2$, are the leading ones. The spin
boundary operator $\mathcal{O}^{(s)}(0)$ is the same operator that drives the
critical behavior in the two-channel spin Kondo problem and is obtained by
contracting the spin-1 field ${\mbox{\boldmath
$\phi$}}^{(s)}(0)$ with the vector of SU(2)$_{2}$ raising operators
$\mbox{\boldmath $J$}_{-1}^{(s)}:\mathcal{O}^{(s)}%
(0)=\mbox{\boldmath $J$}_{-1}^{(s)}\cdot{\mbox{\boldmath $\phi$}}^{(s)}(0)$.
The flavor boundary operator $\lambda_{f}\mathcal{O}^{(f)}(0)$ has the same
structure. In obvious notation: $\mathcal{O}^{(f)}%
(0)=\mbox{\boldmath $J$}_{-1}^{(f)}\cdot{\mbox{\boldmath $\phi$}}^{(f)}(0)$.

In the case of the two-channel (spin) Kondo problem the flavor operator is
effectively suppressed \cite{ALNP}: Of the two available energy scales, the
\emph{bandwidth} $D$ and the \emph{Kondo temperature} $T_{K}$ (where $T_{K}$
sets the scale for the crossover from weak coupling (high-temperature phase)
to strong renormalized coupling (low-temperature phase)), only $D$ enters the
expression for the flavor scaling field $\lambda_{f}$. This is so since the
Kondo temperature is dynamically generated in the spin sector (as indicated by
the infrared divergences in perturbation theory) and hence cannot influence
the scaling of the flavor degrees of freedom. On dimensional grounds one
concludes that $\lambda_{f}\sim\mathcal{O}(1/\sqrt{D})$, whereas $\lambda
_{s}\sim\mathcal{O}(1/\sqrt{T_{{\footnotesize K}}})$. For a small Kondo
coupling, call it $\lambda$, $T_{{\footnotesize K}}\sim
D\,\mbox{exp}(-1/\lambda)\ll D$, and the critical behavior is therefore driven
by $\mathcal{O}^{(s)}(0)$ alone. As we showed in Ref.~[\onlinecite{JAB}], the
picture for the two-channel Anderson model is more involved. There are here
\emph{two} dynamically generated temperature scales $T_{s,f}(\epsilon)$, both
parametrized by $\epsilon=\epsilon_{q}-\epsilon_{s}$ and thus varying with the
position on the fixed line via the dependence of the impurity valence $n_{c}$
on $\epsilon$ \cite{BolechAndrei}. The scaling fields $\lambda_{s,f}$ are
parametrized accordingly \cite{JAB}:
\begin{equation}
\lambda_{s,f}=\frac{B_{s,f}}{\sqrt{T_{s,f}}}, \label{tempscales}%
\end{equation}
with $B_{s,f}$ dimensionless constants. The precise dependence of
$\lambda_{s,f}$ on $\epsilon$ was extracted in Ref.~[\onlinecite{JAB}] from a
fit to the numerical solution of the thermodynamic \emph{Bethe Ansatz} (TBA)
equations of the model \cite{BolechAndrei}. In the magnetic moment regime
where $\epsilon\ll\mu-\Gamma$ [two-channel (spin) Kondo limit] $T_{s}\ll
T_{f}$ and $\lambda_{s}$ dominates. As $\epsilon$ increases $T_{s}$ and
$T_{f}$ approach each other and become equal when $\epsilon=\mu$ (maximal
mixed valence with no moment formation). Continuing along the fixed line, the
two scales trade places, and eventually, at the quadrupolar critical end point
$(\epsilon\gg\mu-\Gamma)$ one finds that $T_{s}\gg T_{f}$. It follows that in
the two-channel Anderson model \emph{both} boundary operators $\mathcal{O}%
^{(s)}(0)$ and $\mathcal{O}^{(f)}(0)$ come into play, with their relative
importance changing continuously as one moves along the fixed line.

Going back to the scaling Hamiltonian in Eq.~(\ref{ScalingHamiltonian}), the
effect of the exactly marginal term $\lambda_{c}J(0)$ is easily obtained via
the observation that it samples the local charge at the impurity site, with
the scaling field $\lambda_{c}=-n_{c}/4$ measuring the impurity valence per
spin and flavor degree of freedom \cite{JAB}. By the Friedel-Langreth sum rule
in Eq.~(\ref{FriedelLangreth}), the resulting shift of the charge content of
the critical bulk Hamiltonian $H^{\ast}$, $Q\rightarrow Q-n_{c}$, shows up as
a phase shift $\delta_{F}=\pi n_{c}/4$ on the electrons that scatter off the
impurity charge potential at $r=0$. In other words, $\psi_{L\alpha\sigma
}\rightarrow\exp(-i\pi n_{c}/4)\psi_{L\alpha\sigma}$ and $\psi_{R\alpha\sigma
}\rightarrow\exp(i\pi n_{c}/4)\psi_{R\alpha\sigma}$, implying that the
left-right propagators $G_{LR}=G_{RL}^{\ast}$ get phase shifted by
$2\delta_{F}=\pi n_{c}/2$, as indicated in Eq.~(\ref{nonchiralGF}).

To probe the effects from the spin and flavor boundary operators in
Eq.~(\ref{ScalingHamiltonian}) requires a perturbative approach. Passing to a
Lagrangian formalism, the correction $\delta S$ to the Euclidean fixed point
action due to $\mathcal{O}^{(f)}(0)$ and $\mathcal{O}^{(s)}(0)$ in
Eq.~(\ref{ScalingHamiltonian}) can be written as
\begin{equation}
\delta S=\sum_{k=f,s}\lambda_{k}\int_{0}^{\beta}d\tau^{\prime}%
\mbox{\boldmath $J$}_{-1}^{(k)}\cdot{\mbox{\boldmath $\phi$}}^{(k)}%
(\tau^{\prime},0), \label{Scorrection}%
\end{equation}
with $\beta=1/k_{B}T$. To leading order in a perturbative expansion this leads
to the following correction for the left-right propagator:
\begin{multline}
\delta_{\alpha\gamma}\delta_{\sigma\mu}\delta G(\tau;r_{1},r_{2})=\sum
_{k=f,s}\lambda_{k}\int_{0}^{\beta}d\tau^{\prime}\\
\times\left\langle \psi_{L\alpha\sigma}^{\mathop{\phantom{\dagger}}}%
(\tau,r_{1})\mbox{\boldmath $J$}_{-1}^{(k)}\cdot{\mbox{\boldmath $\phi$}}%
^{(k)}(\tau^{\prime},0)\psi_{R\gamma\mu}^{\dagger}(0,r_{2})\right\rangle _{T}.
\label{LRcorr}%
\end{multline}
The index $T$ that appears in Eq.~(\ref{LRcorr}) refers to the
\textquotedblleft finite-\textit{T} geometry\textquotedblright\ $\Gamma
^{+}=\{w=\tau+ir\}$, connected to the half-plane $\mathbb{C}^{+}%
=\{\mbox{Im}z>0\}$ used at zero temperature via the conformal mapping
$w=\left(  \beta/\pi\right)  \arctan\left(  z\right)  $.

The three-point functions in Eq. (\ref{LRcorr}) are completely determined by
conformal invariance up to multiplicative constants $N_{f}$ and $N_{s}$
:\cite{GF1}
\begin{multline}
\delta G_{LR}(\tau,r_{1},r_{2})=i(\lambda_{f}N_{f}+\lambda_{s}N_{s}%
)\,\mbox{e}^{2i\delta_{F}}\left(  \frac{\pi}{\beta}\right)  ^{\frac{7}{2}%
}\label{LRintegral}\\
\times\int_{0}^{\beta}d\tau^{\prime}\frac{[-i\sin\frac{\pi}{\beta}%
(\tau+i(r_{1}+r_{2}))]^{\frac{3}{2}}}{[\sin\frac{\pi}{\beta}(\tau^{\prime
}-\tau-ir_{1})\,\sin\frac{\pi}{\beta}(\tau^{\prime}+ir_{2})]^{\frac{5}{2}}}.
\end{multline}
In the $n_{c}=0$ (quadrupolar) limit where $\lambda_{s}\rightarrow0$ and
$\delta_{{\footnotesize F}}=0$ \cite{JAB}, the theory is invariant under
charge conjugation (particle-hole symmetry). Adapting an argument from
Ref.~[\onlinecite{GF1}] we may use this property to determine the phase of
$N_{f}$, and together with an explicit calculation of $\mid\!N_{f}\!\mid^{2}$
one finds the value
\begin{equation}
N_{f}=3/\sqrt{8}. \label{Nf}%
\end{equation}
The value of $N_{s}^{2}$ is fixed via Eq.~(\ref{Nf}) by the vanishing of the
four-point function \cite{GF1}
\begin{multline}
\!\!\!\!\!\!\!\langle\psi_{L\alpha\sigma}^{\dagger}(\tau_{1}+ir_{1}%
)\psi_{R\alpha\sigma}^{\mathop{\phantom{\dagger}}}(\tau_{1}-ir_{1}%
)\psi_{L\gamma\mu}^{\dagger}(\tau_{2}+ir_{2})\psi_{R\gamma\mu}%
^{\mathop{\phantom{\dagger}}}(\tau_{2}-ir_{2})\rangle\label{vanishing}\\
=(N_{f}^{2}+N_{s}^{2})(r_{1}r_{2})^{1/2}\frac{32}{9\mid\tau_{1}-\tau_{2}%
\mid^{3}},
\end{multline}
and one concludes that
\begin{equation}
N_{s}=-i\frac{3}{\sqrt{8}}, \label{Ns}%
\end{equation}
with the negative sign in Eq.~({\ref{Ns}) following from the condition that
the expression for $\delta G_{LR}$ in Eq.~(\ref{LRintegral}) collapses to that
for the two-channel Kondo model \cite{GF1} in the $n_{c}\rightarrow1$ limit.
The scaling fields $\lambda_{f}$ and $\lambda_{s}$ are the same as those that
parametrize the thermodynamics and can thus be fitted to the exact TBA
solution of the model \cite{JAB} (see the next section). With this fit $\delta
G_{LR}$ in Eq.~(\ref{LRintegral}) will be completely specified. Note that by
time reversal invariance, $\delta G_{RL}=\delta G_{LR}^{\ast}$. }

Turning to the chiral propagators in {Eq.~}(\ref{chiralGF}), it is easy to
verify that the corrections to these from $\delta S$ vanish identically:
$\delta G_{LL}\ (\delta G_{RR})$ is also given by the integral expression in
{Eq.~}(\ref{LRintegral}) but with $r_{2}\rightarrow-r_{2}\ (r_{1}%
\rightarrow-r_{1})$. All zeros of the denominator are located in the upper
(lower) half plane, and the integration contour can be deformed to
Im$\tau\rightarrow-\infty\ (+\infty)$ without crossing any singularity; hence
$\delta G_{LL}=\delta G_{RR}=0$.

The integral expression for $\delta G_{LR}=\delta G_{RL}^{\ast}$ in
{Eq.~}(\ref{LRintegral}) differs from that for the two-channel (spin) Kondo
model in Ref.~[\onlinecite{GF1}] only by having a prefactor $(\lambda
_{f}-i\lambda_{s})\mbox{e}^{i\pi n_{c}/2}$ instead of a single scaling field
$\lambda_{s}\ (\equiv\lambda$ in Ref.~[\onlinecite{GF1}]). (It follows
trivially that the result for the chiral propagators is the same for the two
models.) From this point on we can therefore carry over the analysis intact
from Ref.~[\onlinecite{GF1}], at the end simply taking $\lambda\rightarrow
(\lambda_{f}-i\lambda_{s})\mbox{e}^{i\pi n_{c}/2}$. This gives for the
finite-temperature and frequency self-energy\cite{GF1},
\begin{multline}
\Sigma(\omega_{n})=\frac{n_{i}\,\mbox{sgn}(\omega_{n})}{2\pi
ig_{{\footnotesize F}}}\left\{  \vphantom{\int}\right.  1-\frac{3}{\sqrt{2}%
}\left(  \lambda_{f}-i\lambda_{s}\right)  \,\mbox{e}^{i\frac{\pi}{2}n_{c}%
}\left(  \frac{2\pi}{\beta}\right)  ^{\frac{1}{2}}\label{correctedSigma}\\
\qquad\qquad\times\int_{0}^{1}du\left[  u^{\beta\,\mid\,\omega_{n}\,\mid/2\pi
}u^{-1/2}\left(  1-u\right)  ^{1/2}F(u)\right. \\
-\left.  \Gamma^{-2}\left(  3/2\right)  u^{-1/2}\left(  1-u\right)
^{-3/2}\right]  \left.  \vphantom{\int}\right\}  \text{,}%
\end{multline}
where $F\left(  u\right)  $ is the hypergeometric function $F\left(
3/2,3/2,1;u\right)  $.

To summarize this section: to leading order in temperature and frequency, the
exact single-electron Matsubara Green's function $\mathcal{G}\left(
\omega_{n},k\right)  $ of the two-channel Anderson model for a dilute
distribution of impurities is given by Eq.~(\ref{3DGF}) with the self-energy
$\Sigma\left(  \omega_{n}\right)  $ in Eq.~(\ref{correctedSigma}). These are
bulk-electron quantities; in the next section we make an aside to discuss
their connection with the impurity response.

\section{Impurity Self-Energy}

For the sake of simplicity, we will carry out this discussion at zero
temperature. By analytic continuation to real frequencies,
\begin{equation}
\lim_{i\omega_{n}\rightarrow\omega+i0^{+}}\Sigma(\omega_{n})=\Sigma^{R}%
(\omega)\text{,} \label{retardedSE}%
\end{equation}
one obtains from Eq. (\ref{correctedSigma}) an integral expression for the
retarded electron self-energy $\Sigma^{R}(\omega)$. By taking the
$T\rightarrow0$ limit and approximating the integral as done in
Ref.~[\onlinecite{GF1}] one finds that \cite{footnote3}
\begin{align}
\Sigma_{T=0}^{R}\left(  \omega\right)   &  =\frac{n_{i}}{2\pi
ig_{{\footnotesize F}}}\times\label{SelfRetarded}\\
&  \times\left[  1+\frac{12}{\sqrt{\pi}}\left(  \mathcal{A}_{1}+i\mathcal{A}%
_{2}\right)  \,\left(  1-i\operatorname{sgn}\left(  \omega\right)  \right)
\,\left\vert \omega\right\vert ^{\frac{1}{2}}\right] \nonumber
\end{align}
with
\begin{equation}
\left\{
\begin{array}
[c]{c}%
\mathcal{A}_{1}\left(  \lambda_{f},\lambda_{s},n_{c}\right)  \equiv\lambda
_{f}\cos\left(  n_{c}\pi/2\right)  +\lambda_{s}\sin\left(  n_{c}\pi/2\right)
\\
\mathcal{A}_{2}\left(  \lambda_{f},\lambda_{s},n_{c}\right)  \equiv\lambda
_{f}\sin\left(  n_{c}\pi/2\right)  -\lambda_{s}\cos\left(  n_{c}\pi/2\right)
\end{array}
\right.  .
\end{equation}
Writing down the relevant equations of motion, one can make a connection
between the self-energy in Eq. (\ref{SelfRetarded}) and the impurity Green's
function \cite{Bickers}. Defining the latter one following the same
conventions as in Ref.~[\onlinecite{Anders}], the relation reads
\begin{equation}
\Sigma^{R}\left(  \omega\right)  =n_{i}\frac{V^{2}}{2\pi g_{{\footnotesize F}%
}}G_{{\footnotesize imp}}^{R}\ \text{.}%
\end{equation}
Parametrizing the impurity Green's function with a spectral weight equal to
$1/2$, a hybridization amplitude \cite{JAB} $\Gamma\equiv\pi\rho V^{2}$, and a
self-energy $\Sigma_{{\footnotesize imp}}^{R}\left(  \omega\right)  $, we can
extract
\begin{equation}
\Sigma_{{\footnotesize imp}}^{R}\left(  \omega\right)  =\omega-\epsilon
+i\Gamma\left(  1-\frac{n_{i}}{2\pi ig_{{\footnotesize F}}}\left[  \Sigma
^{R}\left(  \omega\right)  \right]  ^{-1}\right)  \ \text{.}%
\end{equation}
The resulting formula for the impurity self-energy inherits from $\Sigma
^{R}\left(  \omega\right)  $ a range of validity for $\left\vert
\omega\right\vert \ll T_{{\footnotesize K}}$, where $T_{{\footnotesize K}%
}\equiv4\Gamma/\pi^{2}\ e^{-\frac{\pi}{2}\left\vert \epsilon\right\vert
/\Gamma}$ is the two-channel Kondo temperature \cite{BolechAndrei}.%

\begin{figure}
[t]
\begin{center}
\includegraphics[
height=2.3333in,
width=3.3278in
]%
{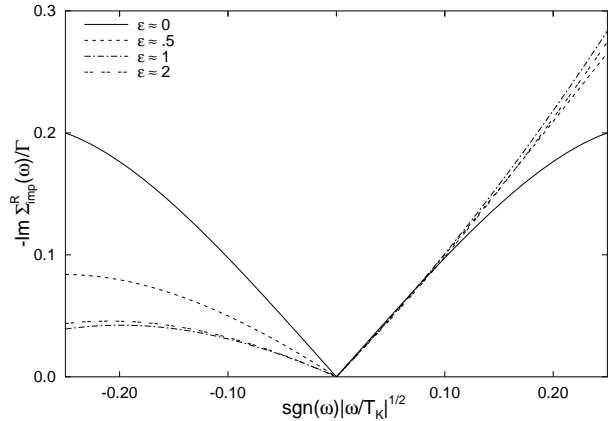}%
\caption{Imaginary part of the zero-temperature impurity self-energy as a
function of frequency. Different curves correspond to different values of the
impurity configurational energy splitting ($\epsilon$). The frequencies are
scaled with the Kondo temperature $T_{{\footnotesize K}}$.}%
\label{fig:selfenergy}%
\end{center}
\end{figure}

In Fig.~\ref{fig:selfenergy} we show the imaginary part of the retarded
impurity self-energy as a function of frequency (scaled with the Kondo
temperature) for different values of the energy splitting $\epsilon$. Only the
curves for positive values of this parameter are shown in the figure, since
for negative values one simply has to use the relation $\operatorname{Im}%
\Sigma_{{\footnotesize imp}}^{R}\left(  \omega,-\epsilon\right)
=\operatorname{Im}\Sigma_{{\footnotesize imp}}^{R}\left(  -\omega
,\epsilon\right)  $ to obtain the corresponding curves. These results are in
fair agreement with the ones obtained recently using the Numerical
Renormalization Group (NRG) mrthod \cite{Anders}. Additionally, notice that
the figure required knowledge of $\mathcal{A}_{1}$ and $\mathcal{A}_{2}$ as
functions of $\varepsilon$; we shall comment on this in the next section.

\section{Resistivity}

Given the retarded electron self-energy $\Sigma^{R}(\omega)$, defined by Eqs.
(\ref{correctedSigma}) and (\ref{retardedSE}), one can readily obtain the
resistivity $\rho(T)$ of the model to leading order in temperature. Adapting
the argument in Appendix C of Ref.~[\onlinecite{GF1}], the assumption that the
impurity-electron interaction is well described using the two-channel $s$-wave
decomposition in Eq.~(\ref{swaveprojection}) implies that the resistivity can
be expressed directly in terms of Im$\Sigma^{R}(\omega)$. From
Eqs.~(\ref{correctedSigma}) and (\ref{retardedSE}) one obtains
\begin{multline}
\mbox{Im}\Sigma^{R}(\omega)=-\frac{n_{i}}{2\pi g_{{\footnotesize F}}}\left\{
\vphantom{\int}\right.  1-3\left(  \frac{\pi}{\beta}\right)  ^{\frac{1}{2}%
}\left[  \vphantom{\int}\right.  \mathcal{A}_{1}(\lambda_{f},\lambda_{s}%
,n_{c})\times\\
\times\int_{0}^{1}du\left[  \cos(\beta\omega(\ln u)/2\pi)u^{-\frac{1}{2}%
}(1-u)^{\frac{1}{2}}F(u)\right. \\
\qquad\qquad-\left.  \Gamma^{-2}(3/2)u^{-\frac{1}{2}}(1-u)^{-\frac{3}{2}%
}\right] \\
\qquad\qquad\qquad\qquad\qquad\qquad\qquad\qquad+~\mathcal{A}_{2}(\lambda
_{f},\lambda_{s},n_{c})\times\\
\qquad\times\int_{0}^{1}du~\sin\left(  \beta\omega\left(  \ln u\right)
/2\pi\right)  u^{-\frac{1}{2}}\left(  1-u\right)  ^{\frac{1}{2}}F\left(
u\right)  \left.  \vphantom{\int}\right]  \!\!\!\left.
\vphantom{\int}\right\}  , \label{ImaginaryRetarded}%
\end{multline}
We have here used the property that $u^{\beta\,\mid\,\omega_{n}\,\mid/2\pi
}\mbox{sgn}(\omega_{n})\rightarrow\cos\left(  \frac{\beta\omega}{2\pi}\ln
u\right)  -i\sin\left(  \frac{\beta\omega}{2\pi}\ln u\right)  $ under the
analytic continuation $i\omega_{n}\rightarrow\omega+i0^{+}$.

As shown in Ref.~[\onlinecite{Bickers}] for this class of problems, vertex
corrections to the resistivity involve $s$-wave correlations which vanish
identically. It follows that the Kubo formula for the resistivity contains
only the quasiparticle lifetime $\tau$, with no weighting over large-angle
scattering processes. With \textit{two} local orbital channels ($\alpha=\pm$)
of charge conduction, the Kubo formula thus reads \cite{Mahan}
\begin{equation}
\rho^{-1}(T)=\frac{4e^{2}}{3m_{e}}\int\frac{d^{3}k}{(2\pi)^{3}}\left[
-\frac{dn_{{\footnotesize F}}(\epsilon_{k})}{d\epsilon_{k}}\right]  k^{2}%
\tau(\epsilon_{k}). \label{inverseResistivity}%
\end{equation}
Here $e$ and $m_{e}$ are the electron charge and mass, respectively,
$n_{{\footnotesize F}}(\epsilon_{k})$ is the Fermi distribution function, and
$\tau(\epsilon_{k})$ is the lifetime of a quasiparticle of energy
$\epsilon_{k}=k^{2}/2m_{e}$,
\begin{equation}
\tau(\epsilon_{k})=-\frac{1}{2}\left[  \mbox{Im}\Sigma^{R}(\epsilon
_{k})\right]  ^{-1}. \label{lifetime}%
\end{equation}
Combining Eqs. (\ref{ImaginaryRetarded}), (\ref{inverseResistivity}), and
(\ref{lifetime}), it follows that
\begin{multline}
\rho^{-1}(T)=\frac{4\pi(eg_{{\footnotesize F}}v_{{\footnotesize F}})^{2}%
}{3n_{i}}\left\{  \vphantom{\int}\right.  1+3\left(  \frac{\pi}{\beta}\right)
^{\frac{1}{2}}\int_{-\infty}^{\infty}\frac{dx}{4\cosh^{2}(x/2)}\\
\qquad\qquad\qquad\qquad\times\left[  \vphantom{\int}\right.  \mathcal{A}%
_{1}(\lambda_{f},\lambda_{s},n_{c})\\
\qquad\times\int_{0}^{1}du\,\left[  \cos(x(\ln u)/2\pi)u^{-\frac{1}{2}%
}(1-u)^{\frac{1}{2}}F(u)\right. \\
\qquad\qquad-\left.  \Gamma^{-2}(3/2)u^{-\frac{1}{2}}(1-u)^{-\frac{3}{2}%
}\right] \\
\qquad\qquad\qquad\qquad+\mathcal{A}_{2}(\lambda_{f},\lambda_{s},n_{c})\\
\times\int_{0}^{1}du\,\sin(x(\ln u)/2\pi)u^{-\frac{1}{2}}(1-u)^{\frac{1}{2}%
}F\left(  u\right)  \left.  \vphantom{\int}\right]  \!\!\!\left.
\vphantom{\int}\right\}  \text{,}%
\end{multline}
where $x\equiv\epsilon_{k}/k_{{\footnotesize B}}T$. We have here used that
$[dn_{{\footnotesize F}}/d\epsilon_{k}]$ in Eq.~(\ref{inverseResistivity})
rapidly goes to zero away from the Fermi level, allowing us to approximate the
momenta that appear in the integral by $k_{{\footnotesize F}}~$($=m_{e}%
v_{{\footnotesize F}}$). (Note that previously $v_{{\footnotesize F}}$ was set
to unity. We still use units where $\hbar=1$.) Carrying out the integrals over
$x$ and subsequently over $u$ (this second integral has to be done
numerically, but to machine accuracy the result is found to be a rational
number and expected to be exact\cite{GF1}), one finally obtains, for the
low-temperature resistivity,
\begin{equation}
\rho\left(  T\right)  =\rho\left(  0\right)  \left[  1+4\left(  \frac{\pi
}{\beta}\right)  ^{\frac{1}{2}}\mathcal{A}_{1}(\lambda_{f},\lambda_{s}%
,n_{c})\right]  \text{,} \label{Resistivity}%
\end{equation}
with
%
\begin{equation}
\rho\left(  0\right)  =\frac{3n_{i}}{4\pi\left(  eg_{{\footnotesize F}%
}v_{{\footnotesize F}}\right)  ^{2}}\text{.}%
\end{equation}
Since both $\lambda_{f}$ and $\lambda_{s}$ on the one hand\cite{JAB} and
$n_{c}$ on the other hand\cite{BolechAndrei} are known to be functions of
$\epsilon$, the expression for the leading low-temperature dependence is
related to the original impurity Hamiltonian via this single parameter.
\begin{figure}
[t]
\begin{center}
\includegraphics[
height=2.3333in,
width=3.3278in
]%
{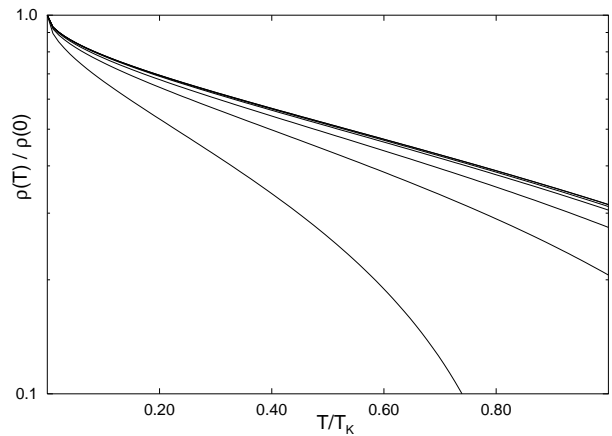}%
\caption{Resistivity \textit{vs} temperature curves for different values of
the microscopic Hamiltonian parameter $\epsilon$. Positive and negative values
of $\epsilon$ fall on top of each other; from left to right the curves
correspond to $\left\vert \epsilon\right\vert \simeq0.0$, $0.5$, $0.8$, $1.1$,
$1.4$, $1.8$, and $2.1$. The curves corresponding to the last two values of
$\epsilon$ fall on top of each other in the scale of the figure, which
illustrates the collapse into a universal curve as the system goes away from
mixed valence.}%
\label{fig:resistivity}%
\end{center}
\end{figure}
This behavior is illustrated in Fig.~\ref{fig:resistivity}. While we have
plotted the curves over the full interval $0\leq T\leq T_{K}$, we should alert
the reader that our results are exact only in the scaling regime $T\ll T_{K}$.
The explicit connection between $\lambda_{f,s}$ and $\epsilon$ was obtained
from a fit of the low-temperature thermodynamics to the results of the TBA
solution \cite{JAB}. But the thermodynamics involves only the squares of the
scaling fields and the sign remains therefore undetermined. However, in the
limit of integer valence we have
\begin{equation}
\lim_{\epsilon/\Gamma\rightarrow\pm\infty}\mathcal{A}_{1}\mathbf{(}\lambda
_{f}\left(  \epsilon\right)  ,\lambda_{s}\left(  \epsilon\right)
,n_{c}\left(  \epsilon\right)  \mathbf{)}=\lambda_{f,s}\left(  \epsilon
\right)  ~,
\end{equation}
and in that limit the model maps onto the weak-coupling two-channel Kondo
model ($J_{{\footnotesize K}}<J^{\ast}$, with $J^{\ast}$ the Kondo fixed point
under renormalization to low energies) for which the sign of $\lambda
_{f,s}\left(  \epsilon\right)  $ is known to be negative (it is expected to
reverse sign for $J_{{\footnotesize K}}>J^{\ast}$) \cite{GF1}. Since the
combined BCFT and TBA analysis indicates that the scaling fields vary
continuously and do not change sign, we conclude that the coefficient
$\mathcal{A}_{1}$ is always negative. Hence, the resistivity is a
monotonically decreasing function of temperature for all values of $\epsilon$.
In the case of $n_{c}$, the Bethe Ansatz solution provides directly an
expression that relates it to $\epsilon$ \cite{BolechAndrei}.

With these considerations the values of $\mathcal{A}_{1,2}(\varepsilon)$ are
completely determined, which allowed us in the previous section to plot the
impurity self-energy. It is interesting to point out that a positive sign for
the scaling fields ($\lambda_{f,s}$) will hamper the comparison of that plot
with the NRG results. Moreover, a positive sign will spoil the causal
properties of the self-energy and is therefore unphysical for the Anderson
model (even in the mixed-valence regime).

Away from $\epsilon\approx0$, we can use Eq.~(\ref{tempscales}) and our
results of Ref.~[\onlinecite{JAB}] to derive an approximate expression for the
resistivity that highlights its scaling properties,
\begin{equation}
\rho\left(  T\right)  /\rho\left(  0\right)  =1-\frac{\sqrt{4\pi}}{3}\left[
\cos\left(  n_{c}\pi/2\right)  +\sin\left(  n_{c}\pi/2\right)  \right]
\sqrt{\frac{T}{T_{{\footnotesize K}}}}.
\end{equation}
Here $T_{{\footnotesize K}}=\min\left\{  T_{f},T_{s}\right\}  $ is the BCFT
Kondo scale and the prefactors are in correspondence with the precise
definition of this scale as given in our previous work. It is important to
remark again that the asymptotic expansion for small temperatures is valid
only for $T\ll T_{{\footnotesize K}}$. For consistency, the same definition of
the Kondo scale \cite{BolechAndrei} was used for both plots in Figs.~1 and 2
(the prefactors in the definition are of course not universal and depend on
the particular convention).

It is here interesting to discuss the experimentally measured low-$T$
resistivity of the thoriated \textrm{UBe}$_{13}$ compound, mentioned in the
Introduction, which shows a $\sqrt{T}$ behavior but with a positive
coefficient. As discussed in Refs.~[\onlinecite{Aliev, Aliev2}], this would
imply that in the (single-impurity) Kondo model framework this system exhibits
a strong electron-impurity coupling ($J_{{\footnotesize K}}>J^{\ast}$). It was
speculated that such a regime was achievable near mixed valence in the context
of the (single-impurity) two-channel Anderson model (see
Ref.~[\onlinecite{CoxZawadowski}] for a review). Our results, however, do not
support those ideas. Perhaps, since \textrm{U}$_{1-x}$\textrm{Th}$_{x}%
$\textrm{Be}$_{13}$ with $x=0.1$ is far from the dilute limit, lattice effects
might play a role in reversing the sign of that coefficient. While later
measurements on the same compound have confirmed the $\sqrt{T}$ behavior with
a positive coefficient \cite{Dickey1}, on the other hand, a $\sqrt{T}$ scaling
of the resistivity with a negative coefficient has been recently observed in a
different uranium-based heavy-fermion material, $\text{Sc}_{1-x}\text{U}%
_{x}\text{Pd}_{3}$ (but only for large dopings, $x\approx0.35$ \cite{Dickey2},
when a single-impurity description of the low-temperature physics is not
always expected).

\bigskip

\section{Summary and Discussion}

In a previous article \cite{JAB} we presented the technical details of the
BCFT solution of the two-channel Anderson model and discussed its asymptotic
low-temperature thermodynamics. In that context we were able to make the
connection with the full solution obtained using the Thermodynamic
Bethe-Ansatz formalism and could explicitly match the scaling fields with the
microscopic parameters of the lattice Hamiltonian. Here we completed the task
by calculating dynamical- and transport-related quantities (the
single-electron Green's function, the electron and impurity self-energies, and
the resistivity). Using our previous results from Ref.~[\onlinecite{JAB}],
these quantities are parametrized directly in terms of the energy difference
between impurity configurations in the original Hamiltonian ($\epsilon$). We
have shown, in particular, how our analytic expression for the impurity
self-energy captures the low-frequency behavior in agreement with the results
of other nonperturbative techniques like Wilson's numerical renormalization
group method \cite{Anders}. As we mentioned in the Introduction, having
reliable access to transport quantities is of crucial importance for the
comparison and interpretation of experiments that continue to seek
indisputable realizations of multi-channel Kondo physics. The work presented
here furthers our understanding of two-channel Kondo physics in mixed-valent
scenarios, thus widening the range of possible candidate systems for
experimental realizations.

\begin{acknowledgments}
We would like to thank F.~B.~Anders for discussions about his recent NRG
results for this model and for useful correspondence. H.J.~acknowledges
support from the Swedish Research Council. C.J.B.~was partially suported by
the MaNEP initiative of the Swiss National Science Foundation.
\end{acknowledgments}


\begin{thebibliography}{99}
\bibitem {Vfootnote}As follows from the decomposition in Eq.
(\ref{swaveprojection}), the hybridization matrix element $V$ in Eq.
(\ref{hybr}) is related to the original 3D matrix element, $V^{\prime}$ call
it, by $V = (k_{F}^{2}/2\pi^{2}) V^{\prime}$.
\bibitem {footnote2}While indices $\alpha$ and $\sigma$ transform according to
the corresponding fundamental representations of $\mathrm{SU}(2)$, the overbar
on the index $\bar{\alpha}$ indicates that it transforms instead according to
the complex conjugate representation.
\bibitem {footnote3}{In the appropriate limits this expression reduces to the
one for the two-channel Kondo model presented in Ref.~[\onlinecite{GF1}].
Note, however, the misprint in the relevant Eqs.~(1.10) and (3.59) of that
reference.}
\end{thebibliography}
\end{document}